\documentstyle[aps,prl,multicol,epsfig,amssymb]{revtex}
\newcommand{\bq}{\begin{equation}}
\newcommand{\eq}{\end{equation}}
\newcommand{\bqa}{\begin{eqnarray}}
\newcommand{\eqa}{\end{eqnarray}}
\newcommand{\nn}{\nonumber \\}

\begin{document}
\draft 
\title{Holon Pairing Instability based on the Bethe-Salpeter Equation obtained from the t-J  Hamiltonians  of  both U(1) and SU(2) Slave-boson Symmetries}

\author{Sung-Sik Lee and Sung-Ho Suck Salk$^a$}
\address{Department of Physics, Pohang University of Science and Technology,\\
Pohang, Kyoungbuk, Korea 790-784\\
$^a$ Korea Institute of Advanced Studies, Seoul 130-012, Korea\\}
\date{\today}

\maketitle

\begin{abstract}
We investigate a possibility of holon pairing for bose condensation based on the Bethe-Salpeter equation obtained from the use of the t-J Hamiltonians of both the U(1) and SU(2) slave-boson symmetries. 
It is shown that the vertex function contributed from ladder diagram series involving holon-holon scattering channel in the Bethe-Salpeter equation leads to a singular behavior at a critical temperature at each hole doping concentration, showing the instability of the normal state against holon pairing. 
We find that the holon pairing instability occurs only in a limited range of hole doping, by showing an "arch" shaped bose condensation line in agreement with observation for high $T_c$ cuprates. 
It is revealed that this is in agreement with a functional integral approach of the slave-boson theories.
\end{abstract}
\begin{multicols}{2}
\newpage

Since the advent of high $T_c$ superconductivity, both the U(1) and SU(2) slave-boson approaches to the t-J Hamiltonian have been proposed to study superconductivity for the two-dimensional systems of copper oxides\cite{KOTLIAR}-\cite{WEN}.
Unlike the U(1) slave-boson mean field theory, the recently proposed SU(2) theory of Wen and Lee has a merit of treating the low energy phase fluctuations of order parameters\cite{WEN}.
Recently we proposed a theory of bose condensation by paying attention to the holon pairing channel in the U(1) slave-boson theory\cite{GIMM}, in contrast to other earlier studies concerned with the single holon condensation. 
In this paper the phase fluctuations of the spinon pairing and hopping order parameters are taken care of by the SU(2) theory.
In addition, following our earlier study\cite{LEE} the spinon and holon degrees of freedom are introduced into the Heisenberg term on the basis of on-site charge fluctuations which arise as a result of site-to-site electron(and thus holon) hopping in two-dimensional quantum systems of hole-doped high $T_c$ cuprates. 
Most of the slave-boson theories have been concerned with single holon bose condensation.
These theories predicted a linear increase of bose condensation temperature with doping concentration, contrary to the observed bose condensation line of an 'arch' shape which manifests the presence of optimal doping rate.
Most recently Wen and Lee\cite{WEN} questioned whether the single holon or the holon pair bose condensation is favored with the SU(2) slave-boson theory.
In this work we pay attention to holon pair bose condensation by demonstrating a possibility of holon pairing instability from the study of the Bethe-Salpeter equation obtained from the use of the t-J Hamiltonians of both the U(1) and SU(2) slave-boson symmetries. 
The present approach is shown to predict the observed characteristics of the bose condensation line of an arch shape in excellent agreement with a functional integral approach to the slave-boson theories.

We write the t-J Hamiltonian, 
\begin{eqnarray}
H & = & -t\sum_{<i,j>}(c_{i\sigma}^{\dagger}c_{j\sigma} + c.c.) + J\sum_{<i,j>}({\bf S}_{i} \cdot {\bf S}_{j} - \frac{1}{4}n_{i}n_{j}),
\label{eq:tjmodel1}
\end{eqnarray}
where ${\bf S}_{i} \cdot {\bf S}_{j} - \frac{1}{4}n_{i}n_{j} =  -\frac{1}{2} ( c_{i \downarrow}^{\dagger}c_{j \uparrow}^{\dagger}-c_{i \uparrow}^{\dagger}c_{j \downarrow}^{\dagger}) (c_{j \uparrow}c_{i \downarrow}-c_{j \downarrow} c_{i \uparrow})$.
Here ${\bf S}_{i}$ is the electron spin operator at site $i$, ${\bf S}_{i}=\frac{1}{2}c_{i\alpha}^{\dagger} \bbox{\sigma}_{\alpha \beta}c_{i\beta}$ with $\bbox{\sigma}_{\alpha \beta}$, the Pauli spin matrix element and $n_i$, the electron number operator at site $i$, $n_i=c_{i\sigma}^{\dagger}c_{i\sigma}$.
In the slave-boson representation\cite{KOTLIAR}-\cite{WEN} $c_{\sigma}$, the electron annihilation operator of spin $\sigma$ can be written as a composite of spinon and holon operators.
That is, $c_{\sigma}  =  b^\dagger f_{\sigma}$ in the U(1) theory and $c_{\alpha}  = \frac{1}{\sqrt{2}} h^\dagger \psi_{\alpha}$ in the SU(2) theory with $\alpha=1,2$, where $f_{\sigma}$($b$) is the spinon(holon) annihilation operator in the U(1) theory, and $\psi_1=\left( \begin{array}{c} f_1 \\ f_2^\dagger \end{array} \right)$ and  $\psi_2 = \left( \begin{array}{c} f_2 \\ -f_1^\dagger \end{array} \right)$ and $h = \left( \begin{array}{c}  b_{1} \\ b_{2} \end{array} \right)$ are respectively the doublets of spinon and holon annihilation operators in the SU(2) theory.

After Hubbard Stratonovich transformations in association with the direct, exchange and pairing channels and the saddle point approximation, the effective Hamiltonian is decomposed into the spinon sector and the holon sector separately. 
Here to explore the instability of the normal state against holon pairing, we pay attention to the holon sector of the Hamiltonian. 
The holon Hamiltonian is derived to be, for the U(1) slave-boson theory\cite{LEE}, 
\bqa
H_{t-J,U(1)}^{b} & = &  -t \sum_{<i,j>} \chi_{ij} b_{i}^{\dagger} b_{j} + c.c. \nn
& - &  \frac{J}{2} \sum_{<i,j>} |\Delta^f_{ij}|^2 b_{i}^{\dagger} b_{j}^\dagger b_j b_i
- \mu_{0} \sum_{i}  b_{i}^{\dagger}b_{i} ,
\label{eq:u1_model}
\eqa
where $\chi_{ij}=<  f_{j\sigma}^{\dagger}f_{i\sigma} + \frac{4t}{J(1-\delta)^2} b_{j}^{\dagger}b_{i} >$ is the hopping order parameter and $ \Delta^f_{ij} = <f_{j\uparrow}f_{i\downarrow}-f_{j\downarrow}f_{i\uparrow}>$, the spinon pairing order parameter, and for the SU(2) slave-boson theory,
\bqa
\lefteqn{H_{t-J,SU(2)}^{b}  =  -\frac{t}{2} \sum_{<i,j>} h_{i}^{\dagger} U_{ij} h_{j} + c.c. }\nn
&& -   \frac{J}{2} \sum_{<i,j>,\alpha,\beta} |\Delta^f_{ij}|^2 b_{i\alpha}^{\dagger} b_{j\beta}^\dagger b_{j\beta} b_{i\alpha}
- \mu_{0} \sum_{i}  h_{i}^{\dagger}h_{i} ,
\label{eq:su2_model}
\eqa
where 
$ U_{i,j}  =  \left( \begin{array}{cc} \chi_{ij} & - \Delta^f_{ij} \\
			   - \Delta^{f*}_{ij} & -\chi^{*}_{ij} 
		\end{array} \right)$ is the order parameter matrix\cite{WEN} of hopping order, $ \chi_{ij} $ and spinon pairing order, $ \Delta^f_{ij}$ with
$ \chi_{ij} = < f_{j\sigma}^\dagger f_{i\sigma} + \frac{2t}{J(1-\delta)^2} (b_{j1}^\dagger b_{i1} - b_{i2}^\dagger b_{j2}) >$ and $ \Delta^f_{ij} = <f_{j1}f_{i2}-f_{j2}f_{i1}>$.

Introducing a uniform hopping order parameter, $\chi_{ji}=\chi$, and a d-wave spinon pairing order parameters, $ \Delta_{ji}^{f}= \pm \Delta_f$(the sign $+(-)$ is for the ${\bf ij}$ link parallel to $\hat x$ ($\hat y$)), we obtain from Eq.(\ref{eq:u1_model}) the momentum space representation of the Hamiltonian, for the U(1) theory,
\bqa
\lefteqn{H^b_{t-J, U(1)}  =  \sum_{{\bf k}} \epsilon( {\bf k} ) b_{{\bf k}}^\dagger b_{{\bf k}} }\nn
&& + \frac{1}{2N}\sum_{{\bf k}, {\bf k}^{'}, {\bf q}} v( {\bf k} - {\bf k}^{'}) b_{-{\bf k}^{'} + {\bf q}}^\dagger b_{{\bf k}^{'}}^\dagger b_{{\bf k}} b_{-{\bf k} + {\bf q}},
\label{eq:u1_model_momentum}
\eqa
where $\epsilon( {\bf k} ) = -2t\chi \gamma_{{\bf k}} - \mu_0$, the energy dispersion of holon with $\gamma_{{\bf k}}= ( \cos k_x + \cos k_y )$ and
$ v( {\bf k}^{'} - {\bf k} ) = -J|\Delta_f|^2 \gamma_{{\bf k}} $, the momentum space representation of holon-holon interaction.
Likewise we obtain from Eq.(\ref{eq:su2_model}),
\bqa
\lefteqn{ H^b_{t-J, SU(2)}  =  \sum_{{\bf k}} h_{{\bf k}}^\dagger \left( \begin{array}{cc} -t \chi \gamma_{{\bf k}} - \mu_0 & t \Delta_f \varphi_{{\bf k}} \\ t \Delta_f \varphi_{{\bf k}} & t \chi \gamma_{{\bf k}} - \mu_0   \end{array} \right) h_{{\bf k}} } \nn
&& +  \frac{1}{2N}\sum_{{\bf k}, {\bf k}^{'}, {\bf q}, \alpha, \beta } v( {\bf k} - {\bf k}^{'}) b_{-{\bf k}^{'} + {\bf q}, \beta}^\dagger b_{{\bf k}^{'}, \alpha}^\dagger b_{{\bf k}, \alpha} b_{-{\bf k} + {\bf q}, \beta},
\label{eq:su2_model_momentum}
\eqa
where $\alpha$, $\beta$ ($=1,2$) represent two isospin components of SU(2) holon
 and $\varphi_{{\bf k}} = ( \cos k_x - \cos k_y )$.
$N$ is the total number of sites for the two dimensional lattice of interest.

Considering the t-matrix of involving the particle-particle(holon-holon) scattering  channels, we obtain the Bethe-Salpeter equations in the U(1) slave-boson representation, 
\bqa
\lefteqn{< k^{'}, -k^{'}+q | t | k, -k+q >_{U(1)}  =  v( {\bf k}^{'} - {\bf k} ) }\nn
&&-\frac{1}{N\beta} \sum_{k^{''}}  v( {\bf k}^{'} - {\bf k}^{''} ) g(k^{''}) g( -k^{''}+q) \times \nn
&& < k^{''}, -k^{''}+q | t | k, -k+q >_{U(1)},
\label{dyson_u1}
\eqa
where $ g(k) = \int_0^{\beta} d \tau e^{ik_0 \tau} <T_{\tau}[b_{{\bf k}}(\tau) b_{{\bf k}}^\dagger(0)]>$, the holon Matsubara Green's function in the U(1) slave-boson representation, and in the SU(2) slave-boson representation,
\bqa
\lefteqn{< k^{'}, \alpha^{'}; -k^{'}+q, \beta^{'} | t | k,\alpha; -k+q,\beta >_{SU(2)}  } \nn
&& =  v( {\bf k}^{'} - {\bf k} ) \delta_{\alpha^{'},\alpha} \delta_{\beta^{'},\beta} \nn
&& -\frac{1}{N\beta} \sum_{k^{''},\alpha^{''},\beta^{''}}  v( {\bf k}^{'} - {\bf k}^{''} ) g_{\alpha^{'}\alpha^{''}}(k^{''}) g_{\beta^{'}\beta^{''}}( -k^{''}+q) \nn
&& \times < k^{''}, \alpha^{''}; -k^{''}+q, \beta^{''} | t | k,\alpha; -k+q,\beta >_{SU(2)}, 
\label{dyson_su2}
\eqa
where $g_{\alpha \beta}(k) = \int_0^{\beta} d \tau e^{ik_0 \tau} <T_{\tau}[b_{\alpha {\bf k}}(\tau) b_{\beta {\bf k}}^\dagger(0)]>$, the holon Matsubara Green's function in the SU(2) slave-boson representation.
Here  $k \equiv (k_0, {\bf k})$ is the three component vector of the energy-momentum of holon.
$q = (q_0, {\bf q})$ is the three component vector of the total energy-momentum of holon pair.

After summing over the Matsubara frequencies in Eqs.(\ref{dyson_u1}) and (\ref{dyson_su2}), we obtain the following matrix equation for the t-matrix, in the U(1) slave-boson representation,
\bqa
\sum_{ {\bf k}^{''}} \left( \delta_{{\bf k}^{'}, {\bf k}^{''}} - m_{{\bf k}^{'}, {\bf k}^{''}} ( q_0, {\bf q} )  \right) t_{{\bf k}^{''}, {\bf k}}( q_0, {\bf q}) = v( {\bf k}^{'} -  {\bf k} ),
\label{eq:matrix_equation_u1}
\eqa
where $t_{{\bf k}^{'}, {\bf k}}( q_0, {\bf q} ) \equiv < k^{'}, -k^{'}+q | t | k, -k+q >_{U(1)}$,
\bqa
\lefteqn{ m_{{\bf k}^{'}, {\bf k}^{''}} ( q_0, {\bf q} ) \equiv -\frac{1}{N\beta} \sum_{k^{''}_0}  v( {\bf k}^{'} - {\bf k}^{''} ) g(k^{''}) g( -k^{''}+q) } \nn
&& = -\frac{ v( {\bf k}^{'} - {\bf k}^{''} )}{N\beta} \left[ \sum_{k^{''}_0} \frac{1}{ ik_0^{''} - \epsilon({\bf k}^{''}) } \frac{1}{ i(-k_0^{''}+q_0) - \epsilon(-{\bf k}^{''} + {\bf q}) } \right] \nn
&& = \frac{1}{N} v( {\bf k}^{'} -  {\bf k}^{''} ) \frac{n(\epsilon({\bf k}^{''})) + e^{\beta \epsilon(-{\bf k}^{''}+ {\bf q}) } n(\epsilon(-{\bf k}^{''}+{\bf q}))}{ iq_0 - (\epsilon(-{\bf k}^{''}+{\bf q}) + \epsilon({\bf k}^{''})) },
\eqa
and $n(\epsilon) = \frac{1}{e^{\beta \epsilon} - 1}$, the boson distribution function. 
Similarly, we obtain, in the SU(2) slave-boson representation,
\bqa
\lefteqn{\sum_{ {\bf k}^{''}, \alpha^{''}, \beta^{''} } \left( \delta_{{\bf k}^{'}, {\bf k}^{''}} \delta_{\alpha^{'} \alpha^{''}} \delta_{\beta^{'} \beta^{''}} - m^{\alpha^{'} \beta^{'} \alpha^{''} \beta^{''} }_{{\bf k}^{'}, {\bf k}^{''}} ( q_0, {\bf q} )  \right) t^{ \alpha^{''} \beta^{''} \alpha \beta }_{{\bf k}^{''}, {\bf k}}( q_0, {\bf q} ) }\nn
&& = v( {\bf k}^{'} -  {\bf k} ) \delta_{\alpha^{'} \alpha} \delta_{\beta^{'} \beta}, \hspace{4cm}
\label{eq:matrix_equation_su2}
\eqa
where $t^{ \alpha^{'} \beta^{'} \alpha \beta }_{{\bf k}^{'}, {\bf k}}( q_0, {\bf q} ) \equiv$ $< k^{'}, \alpha^{'}; -k^{'}+q, \beta^{'} | t | k,\alpha; -k+q,\beta >_{SU(2)}$ and
$m^{\alpha^{'} \beta^{'} \alpha^{''} \beta^{''} }_{{\bf k}^{'}, {\bf k}^{''}} ( q_0, {\bf q} ) \equiv $ 
$\frac{1}{N} \sum_{\alpha^{'}_1 \beta^{'}_1} v( {\bf k}^{'} -  {\bf k}^{''} ) \frac{n(E_{\alpha^{'}_1}({\bf k}^{''})) + e^{\beta E_{\beta^{'}_1}(-{\bf k}^{''}+{\bf q})} n(E_{\beta^{'}_1}(-{\bf k}^{''}+{\bf q}))}{ iq_0 - (E_{\alpha^{'}_1}({\bf k}^{''}) + E_{\beta^{'}_1}(-{\bf k}^{''}+{\bf q})) } \times $ 
$ U_{\alpha^{'}\alpha^{'}_1}({\bf k}^{''}) U_{\beta^{'}\beta^{'}_1}( -{\bf k}^{''} + {\bf q}) U^\dagger_{\alpha^{'}_1 \alpha^{''}}({\bf k}^{''}) U^\dagger_{\beta^{'}_1 \beta^{''} }(- {\bf k}^{''} + {\bf q})$. 
Here $E_{1}( {\bf k} ) = E_{{\bf k}} - \mu_0$ and $E_{2}( {\bf k} ) = -E_{{\bf k}} - \mu_0$ are the energy dispersions of the upper and lower bands of holons with $E_{{\bf k}} =  t \sqrt{ (\chi \gamma_{{\bf k}})^2 + ( \Delta_f \varphi_{{\bf k}})^2 }$.
$U_{\alpha \beta}({\bf k}) = \left( \begin{array}{cc} u_{{\bf k}} & -v_{{\bf k}} \\ v_{{\bf k}} & u_{{\bf k}} \end{array} \right)$ is the unitary transformation matrix used for the diagonalization of the one-body holon Hamiltonian in Eq.(\ref{eq:su2_model_momentum}) with $ u_{{\bf k}} = \frac{1}{\sqrt{2}}\sqrt{1-\frac{t\chi \gamma_{{\bf k}}}{E_{{\bf k}}}}$ and $ v_{{\bf k}} = \frac{1}{\sqrt{2}}\sqrt{1+\frac{t\chi \gamma_{{\bf k}}}{E_{{\bf k}}}}$.
It is noted that the t-matrices, $t_{{\bf k}^{'}, {\bf k}}( q_0, {\bf q})$ and $t^{ \alpha^{'} \beta^{'} \alpha \beta }_{{\bf k}^{'}, {\bf k}}( q_0, {\bf q} )$ for the U(1) and SU(2) theories respectively are independent of the Matsubara frequencies $k^{'}_0$ and $k_0$, owing to the consideration of instantaneous holon-holon interaction, $v( {\bf k}^{'} -  {\bf k})$. 

As mentioned earlier, the t-matrix elements are obtained from Eqs.(\ref{dyson_u1}) and (\ref{dyson_su2}), both of which involve summation over the Matsubara frequencies.
Using the matrix equations (\ref{eq:matrix_equation_u1}) and (\ref{eq:matrix_equation_su2}), the poles of the t-matrices are searched for as a function of energy $q_0$ with ${\bf q} = 0$, i.e., the zero total momentum of the holon pair. 
The t-matrices are numerically evaluated from the use of Eqs.(\ref{eq:matrix_equation_u1}) and (\ref{eq:matrix_equation_su2}) for each doping rate and temperature.
The hopping and spinon pairing order parameters in Eqs.(\ref{eq:u1_model}) and (\ref{eq:su2_model}) are the saddle point values evaluated from the usual partition functions involving the functional integrals of slave-boson representation\cite{LEE}.
For both the U(1) and SU(2) t-matrix calculations we choose $J/t=0.3$ in Eqs.(\ref{eq:u1_model}) and (\ref{eq:su2_model}).

At high temperatures, computed poles are found to be positive and real, indicating that there exist no bound states.
As temperature is lowered to a critical(onset) value $T_c$, we find that with the U(1) theory one pole changes its sign from positive to negative, indicating that their exists an instability of the normal state against holon pairing at the onset(critical) temperature $T_c$.
Similarly, with the SU(2) slave-boson theory two poles change their signs from positive to negative at $T_c$; these two poles correspond to the particle-particle(holon-holon) scattering channels of $b_1$-$b_1$ and $b_2$-$b_2$ respectively.
In Fig.1. the onset temperature $T_c$(denoted by solid square in Fig.1) for the appearance of such negative pole(s) is predicted to occur only in a limited range of hole doping concentration, by revealing an "arch" shaped feature of bose condensation line in agreement with observation in the phase diagram of high $T_c$ cuprates.
The onset temperature of negative pole(s) coincide(s) surprisingly well with the critical temperature of holon pair condensation obtained from the functional integral approach(denoted by open square in Fig.1) of the slave-boson theories\cite{LEE}.
Holon-holon scattering occurs above the lowest possible single particle energy $\epsilon({\bf k}=0)$.
The negative pole corresponds to the binding energy of the holon pair.
This is analogous to the binding energy of electron pairs which results from the Cooper's two particle problem of electron-electron scattering only above the Fermi energy $\epsilon({\bf k}_f)$, i.e., the Fermi-surface. 

In order to find the symmetry of the holon pairing order parameter we now compute the eigenvectors of the t-matrix whose poles change their signs at a critical(onset) temperature $T_c$; the eigenvalue equations are 
\bqa
\sum_{{\bf k^{'}}} t_{{\bf k}, {\bf k}^{'}}( q_0 =0, {\bf q}= 0 ) W({\bf k^{'}}) = \lambda W({\bf k}),
\label{eq:eigenequation_u1}
\eqa
for the U(1) theory, where $W ({\bf k})$ is the eigenvector and $\lambda$, the eigenvalue and 
\bqa
\sum_{ {\bf k^{'}}, \alpha^{'}, \beta^{'}}  t^{ \alpha \beta \alpha^{'} \beta^{'} }_{{\bf k}, {\bf k}^{'}} ( q_0=0, {\bf q}= 0 )  W_{\alpha^{'} \beta^{'}}({\bf k}^{'}) = \lambda W({\bf k})_{\alpha \beta},
\label{eq:eigenequation_su2}
\eqa
for the SU(2) theory, where $W_{\alpha \beta}({\bf k})$ is the eigenvector concerned with the SU(2) isospin channels $\alpha$ and $\beta$($\alpha = 1, 2$ and $\beta = 1, 2$ ) and $\lambda$, the corresponding eigenvalue.
For the case of U(1) there are $N$ eigenvectors corresponding to $N$ discrete values of momenta for an $N \times N$ reciprocal lattice.
For SU(2), $4 N$ eigenvectors are available in accordance with the isospin channels of holon($b_\alpha$)-holon($b_\beta$) scattering.
We search for the eigenvectors whose eigenvalues diverge in the limit of  $q_0 =0$  and ${\bf q} = 0$ at an onset(critical) temperature $T_c$, that is, the eigenvectors corresponding to the pole of the t-matrix whose sign changes at $T_c$.
We choose $N = 10 \times 10$ for a reciprocal lattice to determine the symmetry of holon pairing at an underdoping rate, $\delta=0.05$. 

The computed eigenvectors for both the U(1) and SU(2) slave-boson approaches yielded good fits to the s-wave symmetry of the form  $\cos k_x + \cos k_y$ in momentum space ; for the U(1) theory  we have 
\bqa
W ({\bf k}) = A ( \cos k_x +  \cos k_y),
\label{eq:u1_eigenvector}
\eqa
where $A$ is the normalization constant to satisfy $\sum_{{\bf k}} | W ({\bf k}) |^2 = 1$, and for the SU(2) theory,
\bqa
W^e_{\alpha \beta}({\bf k}) = A^e ( \cos k_x +  \cos k_y)( \delta_{\alpha,1} \delta_{\beta,1} + \delta_{\alpha,2} \delta_{\beta,2} ),
\label{eq:su2_eigenvector_even}
\eqa
and
\bqa
W^o_{\alpha \beta}({\bf k}) = A^o ( \cos k_x +  \cos k_y)( \delta_{\alpha,1} \delta_{\beta,1} - \delta_{\alpha,2} \delta_{\beta,2} ),
\label{eq:su2_eigenvector_odd}
\eqa
for $b_1$ holon and $b_2$ holon pairing, where  $A^{e(o)}$ is the normalization constant to satisfy $\sum_{{\bf k}, \alpha, \beta} | W^{e(o)}_{\alpha \beta}({\bf k}) |^2 = 1$.
One of the two computed eigenvectors shows a phase difference of even multiples of $\pi$ between the $b_1$ and $b_2$ holon pairing order parameters and is well fitted by Eq.(\ref{eq:su2_eigenvector_even}); the other computed eigenvector is well fitted by Eq.(\ref{eq:su2_eigenvector_odd}), by showing a phase difference of odd multiples of $\pi$.
The divergence of eigenvalues was seen to occur nearly at the same temperature(within a numerical accuracy of $10^{-4} t$).
Fig.2 displays the U(1) result of the s-wave symmetry of the holon pairing order.
In Figs.2(a) and (b), we show the SU(2) results of the s-wave symmetry of opposite phase between the $b_1$ and $b_2$ holon pairing order parameters.
Both results demonstrated excellent fits to the s-wave symmetry form of Eqs.(\ref{eq:u1_eigenvector}) and (\ref{eq:su2_eigenvector_odd}).
Since the computed results which fits Eq.(\ref{eq:su2_eigenvector_even}) are indistinguishable with the ones shown in both Fig.2 and Figs.3 (a) and thus are not displayed.
In short, both the U(1) and SU(2) theories predicted the s-wave symmetry of holon pairing order.
This is equivalent to the d-wave symmetry of hole pairing order, by noting that the hole is a composite of a holon(boson) of spin $0$ with charge $+e$ and a spinon(fermion) of spin $1/2$ with charge $0$ and realizing the s-wave holon pairing in the presence of the d-wave spinon pairing\cite{KOTLIAR}-\cite{WEN}\cite{LEE}.

In the present study, we found that the bose condensation of an arch shape occurs only in a limited range of hole doping concentration, by searching for the onset temperature of the holon pairing instability.
It is found that the predicted holon pair bose condensation temperature is in precise agreements with the functional integral approaches of both the U(1) and the SU(2) slave-boson theories. 
Most importantly, both of these approaches were found to satisfactorily predict the experimentally observed bose condensation (superconducting) temperature as a function of the hole doping rate, by revealing the presence of the optimal doping rate in high $T_c$ cuprates.

One(SHSS) of us acknowledges the generous supports of Korea Ministry of Education(BSRI-98 and 99) and the Center for Molecular Science at Korea Advanced Institute of Science and Technology.
We thank Tae-Hyoung Gimm, Jae-Hyun Uhm and Ki-Suk Kim for helpful discussions.


\begin{minipage}[c]{9cm}
\begin{figure}
\vspace{0cm}
\epsfig{file=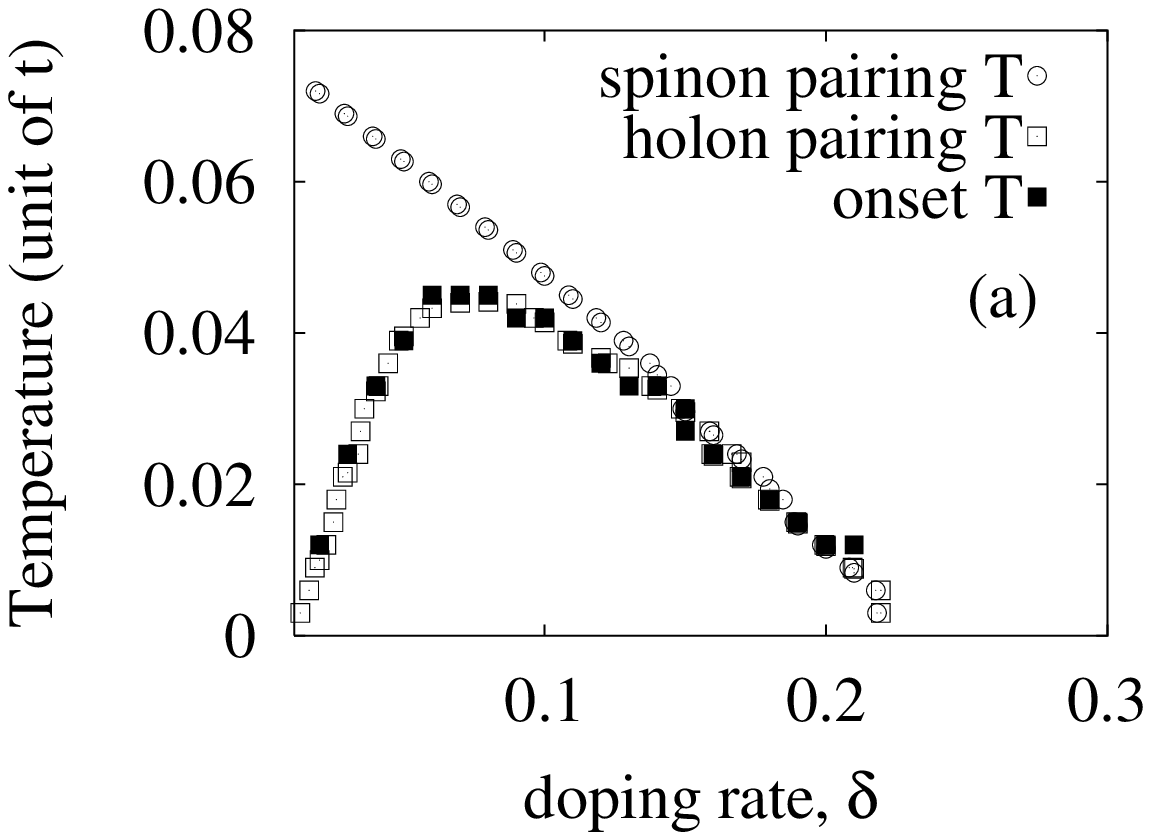,angle=0, height=4.5cm, width=7cm}
\end{figure}
\end{minipage}
\begin{minipage}[c]{9cm}
\begin{figure}
\epsfig{file=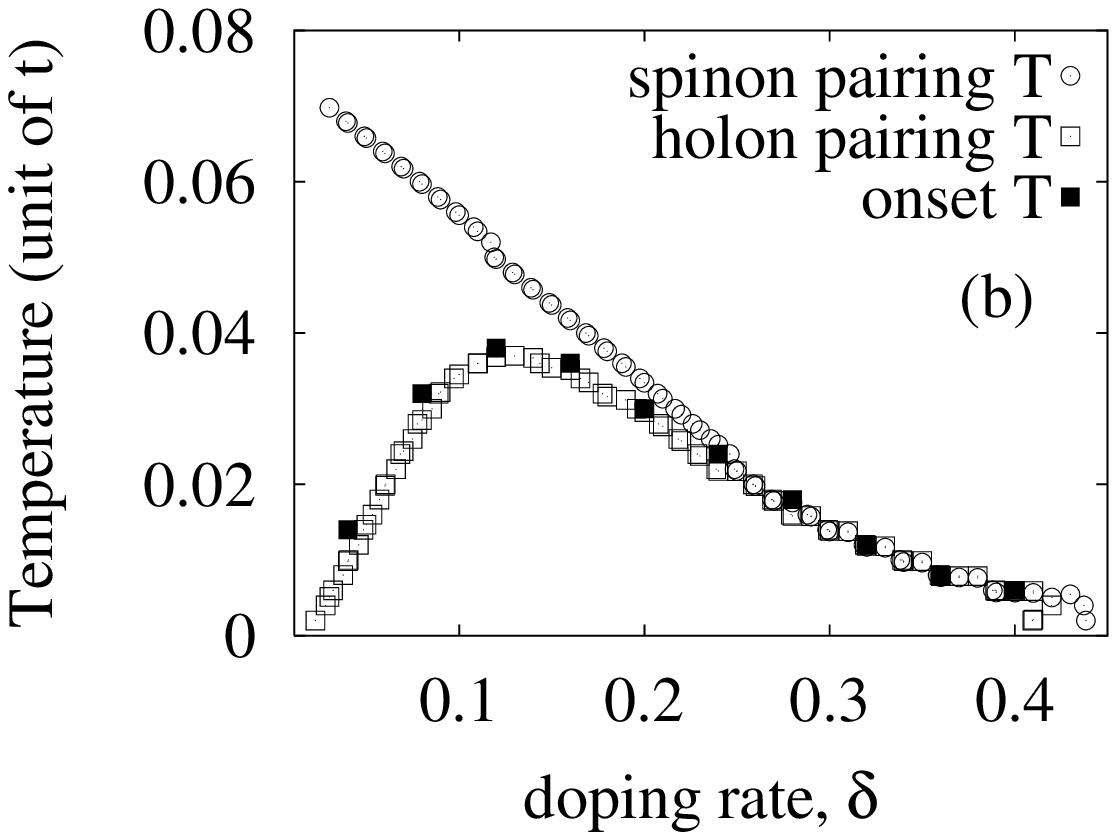,angle=0, height=4.5cm, width=7cm}
\label{fig:1}
\caption{
Onset temperature(denoted by $\blacksquare$) of the negative energy poles representing holon pair bose condensation temperature $T_c$ as a function of hole doping rate for both (a) U(1) and (b) SU(2) slave-boson theories.  The symbols, $\bigcirc$ and $\square$ represent respectively the spinon pairing and holon pairing temperatures obtained from the functional integral approach.
}
\end{figure}
\end{minipage}

\begin{minipage}[c]{9cm}
\begin{figure}
\vspace{0cm}
\epsfig{file=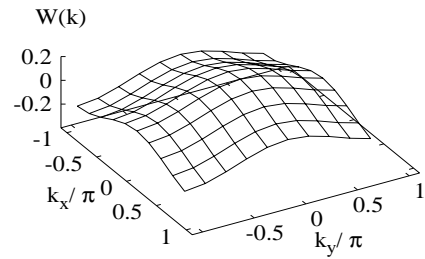,angle=0, height=4.5cm, width=7cm}
\label{fig:2}
\caption{
The computed eigenvector of the U(1) t-matrix which fits the s-wave symmetry of holon pairing in Eq.(13).
}
\end{figure}
\end{minipage}

\begin{minipage}[c]{9cm}
\begin{figure}
\vspace{0cm}
\epsfig{file=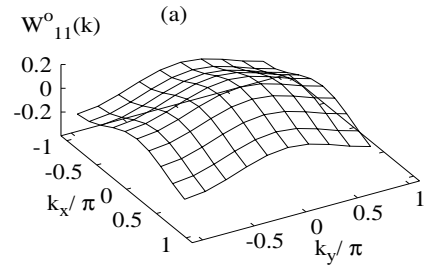,angle=0, height=4.5cm, width=7cm}
\epsfig{file=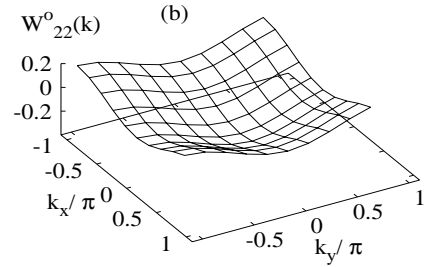,angle=0, height=4.5cm, width=7cm}
\label{fig:3}
\caption{
The computed eigenvector of the SU(2) t-matrix which fits the s-wave symmetry of holon pairing in Eq.(15).
(a) and (b) display the phase difference of $\pi$ between the $b_1$ and $b_2$ holon pair order parameters.
}
\end{figure}
\end{minipage}


\end{multicols}
\end{document}